\documentclass[12pt]{article}
\usepackage[latin1]{inputenc}
\usepackage[british]{babel}
\usepackage{cmap}
\usepackage{lmodern}

\usepackage{amssymb, amsmath, amsthm}
\usepackage[a4paper,top=25mm,bottom=25mm,left=25mm,right=25mm]{geometry}
\usepackage{ragged2e}

\usepackage{authblk} 
\usepackage{pifont}
\usepackage{graphicx}
\usepackage[usenames,dvipsnames,svgnames,table]{xcolor}
\usepackage[figuresright]{rotating}
\usepackage{xtab} 
\usepackage{longtable} 
\usepackage{multirow}
\usepackage{footnote}
\usepackage[stable]{footmisc}
\usepackage{chngpage} 
\usepackage{pdflscape} 
\usepackage[nottoc,notlot,notlof]{tocbibind} 

\usepackage{pgfplots}
\pgfplotsset{compat=1.18}
\pgfplotsset{every tick label/.append style={font=\footnotesize}}
\usepackage{setspace}

\makesavenoteenv{tabular}
\usepackage{tabularx}
\usepackage{booktabs}
\usepackage{threeparttable} 
\usepackage[referable]{threeparttablex} 
\newcolumntype{R}{>{\raggedleft\arraybackslash}X}
\newcolumntype{L}{>{\raggedright\arraybackslash}X}
\newcolumntype{C}{>{\centering\arraybackslash}X}
\newcolumntype{A}{>{\columncolor{gray!25}}C}
\newcolumntype{a}{>{\columncolor{gray!25}}c}

\newlength{\tablen}

\usepackage{dcolumn} 
\newcolumntype{.}{D{.}{.}{-1}}

\usepackage{tikz}
\usetikzlibrary{arrows, calc, matrix, patterns, positioning, trees}
\usepackage[semicolon]{natbib}
\usepackage[hyphens]{url}
\usepackage{hyperref} 
\hypersetup{
  colorlinks   = true,    
  urlcolor     = blue,    
  linkcolor    = blue,    
  citecolor    = ForestGreen      
}
\usepackage{microtype}
\usepackage[singlelinecheck=false,justification=centering]{caption}

\usepackage[labelformat=simple]{subcaption}

\DeclareCaptionLabelFormat{parenthesis}{(#2)}
\makeatletter
\renewcommand\p@subfigure{\arabic{figure}.}
\makeatother

\DeclareCaptionLabelFormat{parenthesis}{(#2)}
\makeatletter
\renewcommand\p@subtable{\arabic{table}.}
\makeatother

\usepackage{enumitem}
\setlist[itemize]{leftmargin=2.5\parindent}
\setlist[enumerate]{leftmargin=2.5\parindent}

\def\addlegendimage{\csname pgfplots@addlegendimage\endcsname}

\theoremstyle{plain}

\theoremstyle{definition}

\theoremstyle{remark}

\def\keywords{\vspace{.5em} 
{\noindent \textit{Keywords}: }}

\def\JEL{\vspace{.5em} 
{\noindent \textbf{\emph{JEL} classification number}: }}

\def\AMS{\vspace{.5em} 
{\noindent \textbf{\emph{MSC} class}: }}

\title{Club coefficients in the UEFA Champions League: \\ Time for shift to an Elo-based formula}
\author{\href{https://sites.google.com/view/laszlocsato}{L\'aszl\'o Csat\'o}\thanks{~E-mail: \emph{laszlo.csato@sztaki.hu}} }
\affil{HUN-REN Institute for Computer Science and Control (HUN-REN SZTAKI) \\
Laboratory on Engineering and Management Intelligence \\
Research Group of Operations Research and Decision Systems}
\affil{Corvinus University of Budapest (BCE) \\
Institute of Operations and Decision Sciences \\
Department of Operations Research and Actuarial Sciences}
\affil{Budapest, Hungary}
\date{\today}

\def\Dedication{
{\noindent
``\emph{Seine Pflicht erkennen und tun, das ist die Hauptsache.}''\footnote{~``\emph{Recognizing and doing one's duty is the main thing.}'' \\
Source: \url{https://www.friedrich-der-grosse.net/zitate-friedrich-des-grossen}}
}
\vspace{0.25cm}

\flushright
\noindent (Frederick the Great)

\vspace{1cm} 
\justify }

\begin{document}


\maketitle
\thispagestyle{empty}
\Dedication

\begin{abstract}
\noindent
One of the most popular club football tournaments, the UEFA Champions League, will see a fundamental reform from the 2024/25 season: the traditional group stage will be replaced by one league where each of the 36 teams plays eight matches. To guarantee that the opponents of the clubs are of the same strength in the new design, it is crucial to forecast the performance of the teams before the tournament as accurately as possible. This paper investigates whether the currently used rating of the teams, the UEFA club coefficient, can be improved by taking the games played in the national leagues into account. According to our logistic regression models, a variant of the Elo method provides a higher accuracy in terms of explanatory power in the Champions League matches. The Union of European Football Associations (UEFA) is encouraged to follow the example of the FIFA World Ranking and reform the calculation of the club coefficients in order to avoid unbalanced schedules in the novel tournament format of the Champions League.


\AMS{62F07, 62P20}

\JEL{C53, Z20}
\end{abstract}

\clearpage
\restoregeometry

\section{Introduction} \label{Sec1}

In team sports, assessing the performance of an individual team is clearly impossible without accounting for the strength of opponents: the same number of goals scored or the same probability of successful attacks are usually more worthy if they have been achieved against a stronger team. This is a crucial issue especially for tournaments that are not played in a round-robin format, for example, if the teams are allocated into different groups, and only the best team(s) from each group qualify for the next phase such as in the FIBA Basketball World Cup or the FIFA World Cup.

Thus, there is a strong demand for reliable measures of teams' strengths. However, it is better to use a transparent methodology for this purpose that can be accepted by all stakeholders, including fans, managers, and players with limited mathematical and statistical knowledge. This implies a trade-off between accuracy and complexity: the calculation of the ratings should be relatively simple without requiring the estimation of difficult regression models even if they would have theoretically a higher predictive power.

The UEFA Champions League, organised by the Union of European Football Associations (UEFA), is the most prestigious club football competition in European football.
UEFA currently uses a particular rating of the teams, the so-called UEFA club coefficient, in order to guarantee the balancedness of the groups in the Champions League and other competitions. This measure is essentially based on the results in the five previous seasons of the UEFA competitions (Champions League, Europa League, Europa Conference League) \citep{UEFA2018g}. 
However, it ignores most matches played by the clubs since it does not depend on the outcomes of matches in the national leagues and cups at all.

Our research question is whether a revised calculation formula for club coefficients can improve performance forecasts in the UEFA Champions League.
In particular, we compare the official UEFA club coefficient and the Football Club Elo Rating, a readily available Elo-based measure of team strength. Their explanatory powers are studied via logistic regression models. Success is defined by winning group matches, winning in the knockout stage, and obtaining a higher rank in the group stage. The database contains the 19 Champions League seasons played between 2003/04 and 2021/22.

The main contribution of the current paper resides in comparing a simple alternative rating of the teams with the widely used UEFA club coefficient with respect to predicting future performance in the UEFA Champions League. According to our knowledge, that has never been done before in the literature.


The study is structured as follows.
A concise overview of literature is provided in Section~\ref{Sec2}. Section~\ref{Sec3} describes the data underlying our empirical investigation, as well as the statistical methods used. The estimations are presented in Section~\ref{Sec4}. Section~\ref{Sec5} discusses the main contributions and outlines some directions for future research. 

\section{Related literature} \label{Sec2}

Our work is connected to at least three directions of academic research.

Some previous papers deal with similar topics in the UEFA Champions League.
\citet{SchokkaertSwinnen2016} empirically examine how the changes in the format of the Champions League have affected uncertainty of outcome. They conclude that qualification in the early rounds has become more predictable but the later stages have become less predictable after 1999.
According to the regression discontinuity design of \citet{EngistMerkusSchafmeister2021}, the UEFA club coefficient itself does not contribute positively to success in this tournament: seeding did not have any effect on the performance of marginally seeded teams.
\citet{Triguero-RuizAvila-Cano2023} find a notable drop in competitive balance in the group stage of the Champions League over the last two decades. This implies a higher accuracy in predicting performance if one finds a reasonable indicator of strength.

The current study is strongly related to research on rating methods in football, too.
\citet{HvattumArntzen2010} implement and test two Elo-based prediction methods.
\citet{LasekSzlavikBhulai2013} and \citet{GasquezRoyuela2016} demonstrate that the Elo rating system is quite competitive in predicting the matches played by national football teams.
\citet{BakerMcHale2015, BakerMcHale2018} suggest time-varying rating models for the English football league and international football teams, respectively.
\citet{VanEetveldeLey2019} give a survey of the most common ranking methods in football.
\citet{LeyVandeWieleVanEeetvelde2019} compare several existing and novel statistical models assigning one or more strength parameters to a football team.

\citet{SzczecinskiDjebbi2020} and \citet{Szczecinski2022} aim to understand and extend the Elo algorithm.
\citet{CeaDuranGuajardoSureSiebertZamorano2020} and \citet{Kaminski2022} discuss the shortcomings of the previous FIFA World Ranking used until 2018.
\citet{LasekGagolewski2021} apply a popular optimisation heuristic (the gradient descent algorithm) to build interpretable rating systems that can be easily adjusted once new results are observed.
According to \citet{SzczecinskiRoatis2022}, the predictive capacity of the current FIFA World Ranking would considerably improve by incorporating home-field advantage and can be further developed by taking the margin of victory into account.
\citet{GyarmatiOrban-MihalykoMihalykoVathy-Fogarassy2023} evaluate three methods with respect to their performance in ranking European football teams.

Finally, some studies examine the predictive power of alternative indicators.
According to \citet{OLeary2017}, the Yahoo crowd outperformed experts at predicting the outcomes of matches played in the 2014 FIFA World Cup and was competitive with the accuracy of betting odds.
\citet{Peeters2018} show that Transfermarkt valuations provide better forecasts for international football matches than standard predictors such as the FIFA ranking and the Elo rating.

\section{Data and methodology} \label{Sec3}

The UEFA Champions League has been organised in the same format between the season of 2003/04 and 2021/22. The 32 clubs are divided into eight groups of four to play a home-away round-robin contest. In each group, the top two teams qualify for the knockout stage. The knockout ties are played in a two-legged format except for the final, which is played in a predetermined neutral stadium. Consequently, the group stage consists of $8 \times 12 = 96$ games, and the knockout stage consists of 14 clashes (28 games) without the final.

In the group stage, each group contains one team from each of the four pots. The pots are primarily determined by the ranking of the teams based on their UEFA club coefficients. However, the titleholder and the champions of the strongest associations have been assigned to Pot 1 between the 2015/16 and 2017/18 seasons \citep{CoronaForrestTenaWiper2019, DagaevRudyak2019}, and the titleholder, the UEFA Europa League titleholder, as well as the champions of the strongest associations, are assigned to Pot 1 since 2018/19 \citep{Csato2020a}. This allocation rule implies that teams having a similar club coefficient usually do \emph{not} play against each other in the group stage.

The UEFA club coefficient is either the sum of all points won in the previous five seasons of European competitions (UEFA Champions League, UEFA Europa League, UEFA Europa Conference League) or the association coefficient over the same period, whichever is the higher \citep{UEFA2018g}. In the Champions League, the first definition usually gives a higher value, although there are a few exceptions such as the German VFL Wolfsburg in the 2021/22 season, which had $14.5$ points but an association coefficient of $14.714$. These data have been collected from the unofficial, but comprehensive website of \emph{Bert Kassies} \citep{Kassies2023a}.

The Elo method has been developed by a Hungarian-born American physics, \emph{\'Arp\'ad \'El{\H o}}, as an improved chess rating system. It is now widely used in many sports, including association football, as shown by the FIFA World Ranking since 2018 \citep{FIFA2018c}. The method calculates the expected winning probability $W$ before any match according to the formula
\begin{equation} \label{eq_Elo}
W = \frac{1}{1 + 10^{\Delta/s}},
\end{equation}
where $\Delta$ is the difference between the Elo rating of the home and the away team, while $s$ is a scaling parameter.
After the outcome of the match becomes known ($R = 1$ for home win; $R=0.5$ for draw; $R=0$ for away win), the ratings are updated: $E_1 = E_0 + K(R-W)$ with $E_0$ being the old and $E_1$ being the new Elo rating. Parameter $K$ reflects the importance of the match and also controls the speed of convergence; a high $K$ implies a quick convergence but makes the ratings volatile, and a low $K$ provides more stable ratings.
Consequently, a win always increases the Elo rating and a loss certainly decreases it. A draw is favourable for the lower-ranked team. The two teams playing each other exchange points, that is, the sum of their ratings remains the same. 

The Elo approach has many variants, we have used the Football Club Elo Ratings, which are available at \url{http://clubelo.com/}. Its formula applies the values $k=400$ and $K=20$, and accounts for home advantage and the margin of victory \citep{FootballClubEloRatings}. This measure has served as the basis of a recent simulation that analysed a recent change in the qualifying system of the UEFA Champions League \citep{Csato2022b}.

Formula~\eqref{eq_Elo} reveals another advantage of an Elo-based approach over the UEFA club coefficient. In the case of the UEFA club coefficient, all wins (draws) in the group stages of the UEFA Champions League, the UEFA Europa League, and the UEFA Europa Conference League increase the rating by the same amount. Analogously, each round that the clubs reach from the Round of 16 in the UEFA Champions League and the UEFA Europa League means the same number of bonus points.
On the other hand, a win against a team with a higher Elo implies a greater update $K(R-W)$ than a win against a team with a lower Elo since $W$ is greater in the second case. Furthermore, the sum of Elo updates for the two opposing teams is always zero. Consequently, the average Elo rating of a national league increases only if its teams perform better than expected in international competitions such as the UEFA Champions League or the UEFA Europa League. The average Elo ratings of domestic leagues are quite different: for example, this was 1,778 for England (the strongest team Manchester City had 2,013) and 1,303 for Hungary (the strongest team Ferencv\'aros had 1,581) on 30 June 2021.

The Elo ratings are continuously updated once a match is played by the teams. On the other hand, the UEFA club coefficient is calculated at the beginning of each season and does not change during the season. Therefore, in order to ensure the fairness of their comparison, we have fixed Football Club Elo Ratings at the level of 30 June each year when most European football leagues, as well as international competitions, have finished. Even though some matches in the first or the preliminary round of the UEFA Champions League qualification have been played before June 30, these teams are missing from our samples since they have never reached the Champions League group stage.

The performance of the UEFA club coefficients and Football Club Elo Ratings are compared by logistic regression models on the basis of the 19 UEFA Champions League seasons played between 2003/04 and 2021/22.

\begin{table}[t!]
  \centering
  \caption{Descriptive statistics of the UEFA club coefficients \\ and Football Club Elo Ratings}
  \label{Table1}
    \rowcolors{1}{}{gray!20}
    \begin{tabularx}{\textwidth}{l CRR RRR} \toprule
    Set of matches & Rating & Mean  & Median & St.~dev. & Minimum & Maximum \\ \bottomrule
    Group matches & UEFA  & 70.49 & 65.07 & 38.75 & 1.63  & 177.00 \\
    Group matches & Elo   & 1,772.54 & 1,777.82 & 136.61 & 1,297.08 & 2,089.27 \\ \hline
    Knockout matches & UEFA  & 95.77 & 96.45 & 35.03 & 12.21 & 177.00 \\
    Knockout matches & Elo   & 1,870.55 & 1,867.13 & 102.58 & 1,549.96 & 2,089.27 \\ \bottomrule
    \end{tabularx}
\end{table}

The descriptive statistics of the two ratings are summarised in Table~\ref{Table1}, separately for the group stage and the knockout stage of the UEFA Champions League seasons that are included in our database. The teams qualifying for the knockout stage have a higher value on average.

For the dependent variable, three options are investigated:
\begin{itemize}
\item
\emph{Group matches}: $1$ indicates home win and $0$ indicates away win, draws are excluded. The explanatory variable is the strength of the home team minus the strength of the away team. \\
From the $19 \times 8 \times 12 = 1{,}824$ group matches, $1{,}402$ were won by one of the teams, which is the number of observations. In particular, there were $863$ home wins (61.6\%).
\item
\emph{Knockout qualification}: $1$ indicates the qualification of the team hosting the first leg and $0$ indicates the qualification of the team hosting the second leg, finals are excluded. The explanatory variable is the strength of the team hosting the first leg minus the strength of the team hosting the second leg. \\
There are $14$ two-legged clashes in each season, which implies $19 \times 14 = 266$ observations. However, in the 2019/20 season, quarterfinals and semifinals were played in a single-leg format on a neutral field behind closed doors. These six matches are removed from the sample. Among the remaining $260$ observations, $159$ (61.2\%) were won by the team hosting the second leg.
\item
\emph{Group ranking}: $1$ indicates that the team having a higher club coefficient is ranked higher and $0$ indicates that the team having a higher club coefficient is ranked lower (two clubs with the same coefficient never played in the same group). \\
In each group, six comparisons can be made between the four teams, meaning $19 \times 8 \times 6 = 912$ observations. The dependent variable equals $1$ for $686$ observations (75.2\%). 
\end{itemize}

Naturally, the COVID-19 pandemic caused some disruptions. Almost all matches were played behind closed doors in the 2020/21 season, which could have affected home advantage \citep{BrysonDoltonReadeSchreyerSingleton2021}. Furthermore, in the 2019/20 season, neither the UEFA Champions League nor many national leagues were finished by 30 June 2020. Thus, the Elo ratings used for 2020/21 do not contain the results of all matches played in the previous seasons. The 2020/21 season will not be considered in some regressions due to this potential bias.

\begin{table}[t!]
  \centering
  \caption{Descriptive statistics of the samples}
  \label{Table2}
    \rowcolors{1}{}{gray!20}
    \begin{tabularx}{\textwidth}{l cRR RRR} \toprule
    Model & Variable & Mean  & Median & St.~dev. & Minimum & Maximum \\ \bottomrule
    Group matches & $\Delta$ UEFA  & 2.89  & 8.26  & 61.79 & $-$159.45 & 159.45 \\
    Group matches & $\Delta$ Elo   & 9.75  & 11.32 & 216.57 & $-$641.00 & 641.00 \\ \hline
    Knockout qualification & $\Delta$ UEFA  & $-$12.70 & $-$18.42 & 48.16 & $-$140.12 & 128.89 \\
    Knockout qualification & $\Delta$ Elo   & $-$53.22 & $-$56.11 & 141.64 & $-$451.45 & 434.14 \\ \hline
    Group ranking & $\Delta$ UEFA  & 50.84 & 47.00 & 32.91 & $-$34.82 & 159.45 \\
    Group ranking & $\Delta$ Elo   & 158.76 & 148.69 & 140.49 & $-$289.73 & 641.00 \\ \bottomrule
    \end{tabularx}
\end{table}

Since our models contain the difference between the ratings of the two teams for each match as explanatory variable(s), the descriptive statistics of the corresponding independent variables are reported in Table~\ref{Table2}.
For group matches, the average difference between the club coefficients and Elo ratings is close to zero. On the other hand, the mean is negative in the knockout stage because the runners-up are guaranteed to host the first game and the group winners are usually stronger teams in the Round of 16. It is worth noting that the highest difference with respect to Elo ratings occurred in the 2021/22 season when Sheriff Tiraspol defeated Real Madrid in Spain, causing one of the biggest shocks in the history of the UEFA Champions League \citep{OConnor2021}.

The scale of the two measures is not the same, the difference between the Elo ratings is about three-four times higher than the difference between the club coefficients. Nonetheless, the variables will not be standardised since we primarily focus on comparing the performance of the models rather than the interpretation of the estimated parameters.

Some standard metrics will be used to evaluate logistic regression models \citep{Allison2013}.
Naturally, the regression is estimated by maximizing the likelihood function. Denote the value of the likelihood function without predictors by $L_0$, the likelihood of the final model by $L_M$, and the number of observations by $n$. Then Cox \& Snell $R^2$ is
\[
R_{C \& S}^2 = 1 - \left( \frac{L_0}{L_M} \right)^{2/n},
\]
which is a generalisation of the usual $R^2$ for linear regression. However, the upper bound of $R_{C \& S}^2$ is not one but
\[
1 - L_0^{2/n} = 1 - \left[ p^p \left( 1-p \right)^{1-p} \right]^2,
\]
where $p$ is the ratio of the event to be predicted in the sample.
Nagelkerke $R^2$ adjusts $R_{C \& S}^2$ by dividing it with its upper bound in order to get a value between zero and one.

Finally, McFadden $R^2$ is defined as
\[
R_{McF}^2 = 1 - \frac{\ln \left( L_M \right)}{\ln \left( L_0 \right)}.
\]
The idea behind the formula is that $\ln \left( L_0 \right)$ can be regarded as the residual sum of squares in a linear regression.

Another measure to assess the performance of a logistic regression model is the area under the ROC curve. The ROC curve plots the true positive rate (sensitivity) as a function of the false positive rate ($1-$ specificity). The area under the ROC curve is the two-dimensional area below the ROC curve from $(0,0)$ to $(1,1)$. Consequently, the area under ROC is 1 if one has a perfect model, for example, each match is won by the team with a higher UEFA club coefficient. If the outcome is random, the area under ROC equals 0.5. A higher value of this indicator shows a more accurate model.

\section{Results} \label{Sec4}

Section~\ref{Sec41} presents the baseline results for the three samples, and Section~\ref{Sec42} examines some other specifications to verify the robustness of the main findings.

\subsection{Comparing the two measures of team strength} \label{Sec41}

\begin{table}[t!]
  \centering
  \caption{Predictive accuracy of the ratings}
  \label{Table3}
    \rowcolors{1}{}{gray!20}
    \begin{tabularx}{0.6\textwidth}{lCC} \toprule
    Model / Variable & $\Delta$ UEFA & $\Delta$ Elo \\ \bottomrule
    Group matches & 70.68\% & 73.32\% \\
    Knockout qualification & 61.92\% & 65.00\% \\
    Group ranking & 75.22\% & 78.51\% \\ \toprule
    \end{tabularx}
\end{table}

First, in order to motivate the more sophisticated models, the rough prediction accuracy of the UEFA club coefficients and Elo ratings are presented in Table~\ref{Table3} with the assumption that a higher (or equal) coefficient/Elo rating predicts success. As expected, group ranking is the easiest to predict, followed by group matches and knockout qualification. Crucially, the difference between Elo ratings robustly outperforms the difference between club coefficients.

\begin{table}[t!]
  \centering
  \caption{Logistic regression models, group matches, 2003/04--2021/22}
  \label{Table4}
\begin{threeparttable}
    \rowcolors{1}{}{gray!20}
    \begin{tabularx}{0.8\textwidth}{lCCC} \toprule \hiderowcolors
          & \multicolumn{1}{c}{(1)} & (2)   & (3) \\ \midrule
    \multirow{2}[0]{*}{Constant} & 0.559*** & 0.591*** & 0.591*** \\
          & (0.064) & (0.067) & (0.067) \\
    \multirow{2}[0]{*}{UEFA} & 0.019*** & \multirow{2}[0]{*}{---} & 0.003 \\
          & (0.001) &       & (0.002) \\
    \multirow{2}[0]{*}{Elo} & \multirow{2}[0]{*}{---} & 0.007*** & 0.006*** \\
          &       & (0.000) & (0.001) \\ \hline \showrowcolors
    Cox \& Snell $R^2$ & 0.222 & 0.270 & 0.272 \\
    Nagelkerke $R^2$ & 0.301 & 0.367 & 0.369 \\
    Classification & 73.0\% & 75.4\% & 75.2\% \\
    Area under ROC & 0.784 & 0.814 & 0.815 \\
    Observations & 1,402  & 1,402  & 1,402 \\ \toprule
    \end{tabularx}
\begin{tablenotes} \footnotesize
\item
Standard errors are in parentheses. * $p < 5\%$; ** $p < 1\%$; *** $p < 0.1\%$.
\item
UEFA (Elo) is the difference between the UEFA club coefficient (Football Club Elo Rating) of the home team and the away team.
\item
Classification is the probability of cases correctly classified if the cut is at $0.5$.
\end{tablenotes}
\end{threeparttable}
\end{table}

Table~\ref{Table4} shows logistic regressions for group matches won by one of the teams.
The constant is highly significant, playing on the home field means a substantial advantage. Elo rating is a stronger predictor of success than UEFA club coefficient: model (2) has a higher explanatory power than model (1), and Elo rating is significant in model (3), while the club coefficient has no additional value here.

\begin{table}[t!]
  \centering
  \caption{Logistic regression models, knockout qualification, 2003/04--2021/22}
  \label{Table5}
\begin{threeparttable}
    \rowcolors{1}{}{gray!20}
    \begin{tabularx}{0.8\textwidth}{lCCC} \toprule \hiderowcolors
          & \multicolumn{1}{c}{(1)} & (2)   & (3) \\ \midrule
    \multirow{2}[0]{*}{Constant} & $-$0.362** & $-$0.223 & $-$0.218 \\
          & (0.132) & (0.140) & (0.141) \\
    \multirow{2}[0]{*}{UEFA} & 0.009** & \multirow{2}[0]{*}{---} & $-$0.007 \\
          & (0.003) &       & (0.004) \\
    \multirow{2}[0]{*}{Elo} & \multirow{2}[0]{*}{---} & 0.005*** & 0.007*** \\
          &       & (0.001) & (0.002) \\ \hline \showrowcolors
    Cox \& Snell $R^2$ & 0.038 & 0.109 & 0.117 \\
    Nagelkerke $R^2$ & 0.051 & 0.148 & 0.159 \\
    Classification & 59.6\% & 65.4\% & 68.1\% \\
    Area under ROC & 0.617 & 0.690 & 0.693 \\
    Observations & 260   & 260   & 260 \\ \toprule
    \end{tabularx}
\begin{tablenotes} \footnotesize
\item
Standard errors are in parentheses. * $p < 5\%$; ** $p < 1\%$; *** $p < 0.1\%$.
\item
UEFA (Elo) is the difference between the UEFA club coefficient (Football Club Elo Rating) of the team hosting the first leg and the team hosting the second leg.
\item
Classification is the probability of cases correctly classified if the cut is at $0.5$.
\item
In the 2019/20 season, quarterfinals and semifinals were played in a single-leg format on a neutral field behind closed doors. These six matches are removed from the sample.
\end{tablenotes}
\end{threeparttable}
\end{table}

According to Table~\ref{Table5}, the same conclusions hold for the two-legged clashes played in the knockout stage. Again, model (2) outperforms model (1), and using the club coefficients is not able to improve the predictions based on Elo ratings.

\begin{table}[t!]
  \centering
  \caption{Logistic regression models, group ranking, 2003/04--2021/22}
  \label{Table6}
\begin{threeparttable}
    \rowcolors{1}{}{gray!20}
    \begin{tabularx}{0.8\textwidth}{lCCC} \toprule \hiderowcolors
          & \multicolumn{1}{c}{(1)} & (2)   & (3) \\ \midrule
    \multirow{2}[0]{*}{Constant} & $-$0.095 & 0.015 & $-$0.338* \\
          & (0.141) & (0.116) & (0.154) \\
    \multirow{2}[0]{*}{UEFA} & 0.027*** & \multirow{2}[0]{*}{---} & 0.012*** \\
          & (0.003) &       & (0.004) \\
    \multirow{2}[0]{*}{Elo} & \multirow{2}[0]{*}{---} & 0.009*** & 0.008*** \\
          &       & (0.001) & (0.001) \\ \hline \showrowcolors
    Cox \& Snell $R^2$ & 0.107 & 0.172 & 0.184 \\
    Nagelkerke $R^2$ & 0.159 & 0.256 & 0.272 \\
    Classification & 78.2\% & 78.1\% & 78.1\% \\
    Area under ROC & 0.704 & 0.776 & 0.784 \\
    Observations & 912   & 912   & 912 \\ \toprule
    \end{tabularx}
\begin{tablenotes} \footnotesize
\item
Standard errors are in parentheses. * $p < 5\%$; ** $p < 1\%$; *** $p < 0.1\%$.
\item
UEFA (Elo) is the difference between the UEFA club coefficients (Football Club Elo Ratings) of the two teams.
\item
Classification is the probability of cases correctly classified if the cut is at $0.5$.
\end{tablenotes}
\end{threeparttable}
\end{table}

Analogously, Table~\ref{Table6} reveals that the Elo rating is more useful for predicting group ranking than the UEFA club coefficient as model (2) is more efficient than model (1), although the club coefficients now have a significant contribution when both measures of strengths are considered. In particular, a team with a fixed Elo rating is more likely to finish above another team if it has a higher club coefficient. This makes sense as better performance in previous European competitions can provide some experience for the squads that cannot be obtained by playing against domestic teams.

\subsection{Sensitivity analysis} \label{Sec42}

\begin{table}[t!]
  \centering
  \caption{Logistic regression models, group ranking with \\ an alternative dependent variable, 2003/04--2021/22}
  \label{Table7}
\begin{threeparttable}
    \rowcolors{1}{}{gray!20}
    \begin{tabularx}{0.8\textwidth}{lCCC} \toprule \hiderowcolors
          & \multicolumn{1}{c}{(1)} & (2)   & (3) \\ \midrule
    \multirow{2}[0]{*}{Constant} & 0.437** & 0.415*** & 0.286 \\
          & (0.141) & (0.114) & (0.151) \\
    \multirow{2}[0]{*}{UEFA} & 0.019*** & \multirow{2}[0]{*}{---} & 0.005 \\
          & (0.003) &       & (0.004) \\
    \multirow{2}[0]{*}{Elo} & \multirow{2}[0]{*}{---} & 0.007*** & 0.007*** \\
          &       & (0.001) & (0.001) \\ \hline \showrowcolors
    Cox \& Snell $R^2$ & 0.054 & 0.115 & 0.116 \\
    Nagelkerke $R^2$ & 0.083 & 0.177 & 0.180 \\
    Classification & 78.7\% & 78.1\% & 78.5\% \\
    Area under ROC & 0.653 & 0.746 & 0.744 \\
    Observations & 912   & 912   & 912 \\ \toprule
    \end{tabularx}
\begin{tablenotes} \footnotesize
\item
Standard errors are in parentheses. * $p < 5\%$; ** $p < 1\%$; *** $p < 0.1\%$.
\item
UEFA (Elo) is the difference between the UEFA club coefficients (Football Club Elo Ratings) of the two teams.
\item
Classification is the probability of cases correctly classified if the cut is at $0.5$.
\end{tablenotes}
\end{threeparttable}
\end{table}

In the regressions for group ranking (Table~\ref{Table6}), the definition of the dependent variable is somewhat arbitrary as it ``assumes'' that the team with a higher club coefficient should be ranked higher. Therefore, the estimations are repeated with an alternative specification when $1$ ($0$) indicates that the club having a higher Elo is ranked higher (lower). Then the dependent variable equals $1$ for 716 observations (78.5\%).
The results are reported in Table~\ref{Table7}. Again, model (2) has a better fit compared to model (1), but now the UEFA club coefficient is not able to significantly contribute to model (2) as can be seen in model (3).

\begin{table}[t!]
  \centering
  \caption{Multinomial logistic regression models, \\ all group matches including draws, 2003/04--2021/22}
  \label{Table8}
\begin{threeparttable}
    \rowcolors{1}{}{gray!20}
    \begin{tabularx}{1\textwidth}{lLCCC} \toprule \hiderowcolors
          & & \multicolumn{1}{c}{(1)} & (2)   & (3) \\ \midrule
    \multirow{6}[0]{*}{Home win} & \multirow{2}[0]{*}{Constant} & 0.558*** & 0.590*** & 0.590*** \\
          &       & (0.064) & (0.066) & (0.066) \\
          & \multirow{2}[0]{*}{UEFA} & 0.020*** & \multirow{2}[0]{*}{---} & 0.004 \\
          &       & (0.001) &       & (0.002) \\
          & \multirow{2}[0]{*}{Elo} & \multirow{2}[0]{*}{---} & 0.007*** & 0.006*** \\
          &       &       & (0.000) & (0.001) \\ \midrule
    \multirow{6}[0]{*}{Draw} & \multirow{2}[0]{*}{Constant} & $-$0.061 & $-$0.001 & 0.000 \\
          &       & (0.072) & (0.073) & (0.073) \\
          & \multirow{2}[0]{*}{UEFA} & 0.008*** & \multirow{2}[0]{*}{---} & 0.000 \\
          &       & (0.001) &       & (0.002) \\
          & \multirow{2}[0]{*}{Elo} & \multirow{2}[0]{*}{---} & 0.003*** & 0.003*** \\
          &       &       & (0.000) & (0.001) \\ \hline \showrowcolors
    \multicolumn{2}{l}{Cox \& Snell $R^2$} & 0.186 & 0.225 & 0.227 \\
    \multicolumn{2}{l}{Nagelkerke $R^2$} & 0.212 & 0.257 & 0.259 \\
    \multicolumn{2}{l}{McFadden $R^2$} & 0.098 & 0.121 & 0.123 \\
    \multicolumn{2}{l}{Classification} & 56.0\% & 57.9\% & 57.7\% \\
    Area under ROC & Home win & 0.737 & 0.761 & 0.762 \\
    Area under ROC & Draw  & 0.569 & 0.578 & 0.577 \\
    Area under ROC & Away win & 0.734 & 0.761 & 0.761 \\
    \multicolumn{2}{l}{Observations} & 1,824  & 1,824  & 1,824 \\ \bottomrule
    \end{tabularx}
\begin{tablenotes} \footnotesize
\item
Standard errors are in parentheses. * $p < 5\%$; ** $p < 1\%$; *** $p < 0.1\%$.
\item
UEFA (Elo) is the difference between the UEFA club coefficient (Football Club Elo Rating) of the home team and the away team.
\item
The reference category is away win.
\item
Classification is the probability of cases correctly classified if the cut is at $0.5$.
\end{tablenotes}
\end{threeparttable}
\end{table}

Table~\ref{Table4} has shown the results for all group matches without draws, which may contain an inherent distortion if the drawn games are different from the games won by one of the opposing teams. Hence, Table~\ref{Table8} presents multinomial logistic regressions for the three possible outcomes (home win, draw, away win) with the reference category being away win. The number of observations increases to 1,824 as has been presented in the previous section.
A higher club coefficient or a higher Elo rating significantly increases the probability of winning and playing a draw compared to losing. Model (2) clearly outperforms model (1), and the UEFA club coefficient is again insignificant in model (3), thus, the Elo rating remains a better measure of strength. Now there are three alternative interpretations of the area under the ROC curve, which suggest that draws are the most difficult to forecast.

\begin{table}[t!]
  \centering
  \caption{Logistic regression models, group matches, sample split into two periods}
  \label{Table9}
\begin{threeparttable}
    \rowcolors{1}{gray!20}{}
    \begin{tabularx}{\textwidth}{l CCC CCC} \toprule \hiderowcolors
    Period & \multicolumn{3}{c}{2003/04--2011/12} & \multicolumn{3}{c}{2012/13--2021/22 (w/o 2020/21)} \\
    \multicolumn{1}{l}{Model} & (1)   & (2)   & (3)   & (1)   & (2)   & (3) \\ \midrule
    \multirow{2}[0]{*}{Constant} & 0.631*** & 0.662*** & 0.663*** & 0.532*** & 0.558*** & 0.561*** \\
          & (0.093) & (0.096) & (0.096) & (0.093) & (0.097) & (0.097) \\
    \multirow{2}[0]{*}{UEFA} & 0.019*** & \multirow{2}[0]{*}{---} & 0.002 & 0.019*** & \multirow{2}[0]{*}{---} & 0.005 \\
          & (0.002) &       & (0.003) & (0.002) &       & (0.003) \\
    \multirow{2}[0]{*}{Elo} & \multirow{2}[0]{*}{---} & 0.006*** & 0.006*** & \multirow{2}[0]{*}{---} & 0.007*** & 0.005*** \\
          &       & (0.001) & (0.001) &       & (0.001) & (0.001) \\ \hline \showrowcolors
    Cox \& Snell $R^2$ & 0.187 & 0.234 & 0.234 & 0.243 & 0.287 & 0.290 \\
    Nagelkerke $R^2$ & 0.255 & 0.320 & 0.320 & 0.329 & 0.389 & 0.393 \\
    Classification & 72.5\% & 75.0\% & 75.0\% & 73.3\% & 74.5\% & 74.8\% \\
    Area under ROC & 0.764 & 0.792 & 0.815 & 0.797 & 0.824 & 0.826 \\
    Observations & 652   & 652   & 652   & 674   & 674   & 674 \\ \toprule
    \end{tabularx}
\begin{tablenotes} \footnotesize
\item
Standard errors are in parentheses. * $p < 5\%$; ** $p < 1\%$; *** $p < 0.1\%$.
\item
UEFA (Elo) is the difference between the UEFA club coefficient (Football Club Elo Rating) of the home team and the away team.
\item
Classification is the probability of cases correctly classified if the cut is at $0.5$.
\end{tablenotes}
\end{threeparttable}
\end{table}

In order to identify potential trends and check the robustness of the previous findings, the sample has been cut into two equal parts of 9-9 seasons together with removing 2020/21, which was affected by the Covid-19 pandemic to a great extent (see Section~\ref{Sec3}).
According to Table~\ref{Table9}, there is only a slight difference between the parameters for group matches estimated on the basis of the first and the last nine seasons. Nonetheless, the predictions are somewhat more accurate since 2012, which might imply a worsening competitive balance that is in line with previous research \citep{Triguero-RuizAvila-Cano2023}. On the other hand, the dominance of model (2) over model (1) is obvious, and the UEFA club coefficient still does not provide additional information to Football Club Elo Rating.

\begin{table}[t!]
  \centering
  \caption{Logistic regression models, knockout qualification, sample split into two periods}
  \label{Table10}
\begin{threeparttable}
    \rowcolors{1}{gray!20}{}
    \begin{tabularx}{\textwidth}{l CCC CCC} \toprule \hiderowcolors
    Period & \multicolumn{3}{c}{2003/04--2011/12} & \multicolumn{3}{c}{2012/13--2021/22 (w/o 2020/21)} \\ \midrule
    \multicolumn{1}{l}{Model} & (1)   & (2)   & (3)   & (1)   & (2)   & (3) \\
    \multirow{2}[0]{*}{Constant} & $-$0.364 & $-$0.275 & $-$0.275 & $-$0.280 & $-$0.075 & $-$0.049 \\
          & (0.193) & (0.199) & (0.199) & (0.193) & (0.214) & (0.219) \\
    \multirow{2}[0]{*}{UEFA} & 0.013** & \multirow{2}[0]{*}{---} & 0.002 & 0.007* & \multirow{2}[0]{*}{---} & $-$0.015* \\
          & (0.005) &       & (0.007) & (0.003) &       & (0.006) \\
    \multirow{2}[0]{*}{Elo} & \multirow{2}[0]{*}{---} & 0.005** & 0.005* & \multirow{2}[0]{*}{---} & 0.006*** & 0.011*** \\
          &       & (0.002) & (0.002) &       & (0.002) & (0.003) \\ \hline \showrowcolors
    Cox \& Snell $R^2$ & 0.056 & 0.089 & 0.089 & 0.038 & 0.163 & 0.204 \\
    Nagelkerke $R^2$ & 0.076 & 0.121 & 0.122 & 0.052 & 0.220  & 0.275 \\
    Classification & 63.5\% & 67.5\% & 67.5\% & 58.3\% & 67.5\% & 66.7\% \\
    Area under ROC & 0.649 & 0.665 & 0.666 & 0.608 & 0.748 & 0.763 \\
    Observations & 126   & 126   & 126   & 120   & 120   & 120 \\ \toprule
    \end{tabularx}
\begin{tablenotes} \footnotesize
\item
Standard errors are in parentheses. * $p < 5\%$; ** $p < 1\%$; *** $p < 0.1\%$.
\item
UEFA (Elo) is the difference between the UEFA club coefficient (Football Club Elo Rating) of the team hosting the first leg and the team hosting the second leg.
\item
Classification is the probability of cases correctly classified if the cut is at $0.5$.
\item
In the 2019/20 season, quarterfinals and semifinals were played in a single-leg format on a neutral field behind closed doors. These six matches are removed from the sample.
\end{tablenotes}
\end{threeparttable}
\end{table}

Table~\ref{Table10} focuses on qualification in the knockout stage for the two subsamples. Again, model (2) outperforms model (1), and the metrics of goodness of fit are higher for the recent Champions League seasons if the model contains the Elo rating. As before, there is no reason to favour the UEFA club coefficient over the Elo rating with respect to predicting qualification.

\begin{table}[t!]
  \centering
  \caption{Logistic regression models, group ranking, sample split into two periods}
  \label{Table11}
\begin{threeparttable}
    \rowcolors{1}{gray!20}{}
    \begin{tabularx}{\textwidth}{l CCC CCC} \toprule \hiderowcolors
    Period & \multicolumn{3}{c}{2003/04--2011/12} & \multicolumn{3}{c}{2012/13--2021/22 (w/o 2020/21)} \\ \midrule
    \multicolumn{1}{l}{Model} & (1)   & (2)   & (3)   & (1)   & (2)   & (3) \\
    \multirow{2}[0]{*}{Constant} & $-$0.009 & 0.181 & $-$0.154 & $-$0.195 & $-$0.165 & $-$0.539* \\
          & (0.207) & (0.164) & (0.217) & (0.201) & (0.173) & (0.229) \\
    \multirow{2}[0]{*}{UEFA} & 0.026*** & \multirow{2}[0]{*}{---} & 0.013* & 0.027*** & \multirow{2}[0]{*}{---} & 0.012** \\
          & (0.005) &       & (0.005) & (0.004) &       & (0.005) \\
    \multirow{2}[0]{*}{Elo} & \multirow{2}[0]{*}{---} & 0.007*** & 0.006*** & \multirow{2}[0]{*}{---} & 0.010*** & 0.009*** \\
          &       & (0.001) & (0.001) &       & (0.001) & (0.001) \\ \hline \showrowcolors
    Cox \& Snell $R^2$ & 0.088 & 0.125 & 0.136 & 0.125 & 0.213 & 0.225 \\
    Nagelkerke $R^2$ & 0.130  & 0.184 & 0.204 & 0.184 & 0.314 & 0.332 \\
    Classification & 77.3\% & 76.9\% & 77.3\% & 78.5\% & 78.5\% & 78.7\% \\
    Area under ROC & 0.686 & 0.736 & 0.745 & 0.717 & 0.803 & 0.810 \\
    Observations & 432   & 432   & 432   & 432   & 432   & 432 \\ \toprule
    \end{tabularx}
\begin{tablenotes} \footnotesize
\item
Standard errors are in parentheses. * $p < 5\%$; ** $p < 1\%$; *** $p < 0.1\%$.
\item
UEFA (Elo) is the difference between the UEFA club coefficients (Football Club Elo Ratings) of the two teams.
\item
Classification is the probability of cases correctly classified if the cut is at $0.5$.
\end{tablenotes}
\end{threeparttable}
\end{table}

Finally, Table~\ref{Table11} presents the estimations for group ranking. The main message does not change:
(a) the Champions League has been more predictable between 2012 and 2022 than between 2003 and 2012;
(b) the Elo rating gives more accurate forecasts compared to the UEFA club coefficient.
Furthermore, similar to Table~\ref{Table6}, adding the club coefficient is able to improve the model as its parameter is significant in equation (3).

\section{Discussion} \label{Sec5}

Ensuring the balancedness of competitions is a fundamental issue of tournament design in order to avoid that a strong team has a lower chance to qualify than a weak team merely because of the outcome of the draw \citep{Csato2021a, Guyon2015a, LaprePalazzolo2022, LaprePalazzolo2023}. According to our findings presented above, the Football Club Elo Rating robustly outperforms the currently used UEFA club coefficient in terms of predictive accuracy. This conclusion does not depend on the sample (group matches, knockout qualification, group ranking). Similarly, both pseudo-$R^2$ values and the areas under ROC support the use of Elo ratings instead of club coefficients.

In the following, it will be highlighted why this result is especially important for the UEFA Champions League.

\subsection{A new challenge in the UEFA Champions League: scheduling should account for the strength of the teams} \label{Sec51}

The format of the UEFA Champions League has essentially not changed between the 2003/04 and 2023/24 seasons, even though there have been some reforms in the entry rules \citep{Csato2019c}, in the use of the away goals rule \citep{Bahamonde-BirkeBahamonde-Birke2023, Jost2021}, in the seeding policy \citep{CoronaForrestTenaWiper2019, Csato2020a, DagaevRudyak2019}, as well as in the design of the qualification system \citep{Csato2022b}.

However, UEFA will introduce a fundamentally new competition format in the 2024/25 season. In particular, the 36 teams will compete in one league where each team plays four matches at home and four matches away instead of the previous six matches against three teams, played on a home-and-away basis \citep{UEFA2022a}. The top eight clubs of the league will qualify for the Round of 16, while the teams ranked between the 9th and 24th places will go to the knockout round play-offs to play two-legged clashes for the remaining eight places in the Round of 16. A similar design will be used in the other two European competitions, the UEFA Europa League and the UEFA Europa Conference League.

The novel competition structure is officially called the ``Swiss system'' \citep{UEFA2022a}. The name has been inspired by a non-eliminating tournament format containing a fixed number of rounds that is widely used in chess \citep{Csato2013a, DongRibeiroXuZamoraMaJing2023}. It is usually applied when the high number of participants allows only to play a considerably fewer number of rounds than required by a round-robin contest. In the original Swiss-system, the pairing of players in each round is determined by the results of previous rounds, ensuring that both opponents have an equal or similar score \citep{CsehFuhrlichLenzner2023, FuhrlichCsehLenzner2021}. This is feasible in chess and some other sports, where the matches can be played at a given location or at least in the same city. However, dynamic scheduling is hardly an option in a football tournament played across the continent since the teams and the fans want to know at the moment of the draw the opponents and the field of all games.

Therefore, the schedule of the UEFA Champions League will be determined according to the following rules \citep{UEFA2023}:
\begin{itemize}
\item
The 36 clubs are seeded into four pots of nine clubs based on their individual club coefficients established at the beginning of the season;
\item
The first pot contains the Champions League titleholder and the top eight clubs in the club coefficient ranking;
\item
Pots 2, 3, and 4 consist of all other clubs according to their ranking order;
\item
Each club is drawn against two opponents from each of the four pots, playing one match at home and one match away against them;
\item
Clubs from the same association are generally not drawn against each other in the league phase;
\item
Exceptionally, one match per club against another club from the same association may be allowed for associations with four or more clubs in the league phase, if this is necessary to avoid a deadlock in the draw.
\end{itemize}

Obviously, it is crucial in this system to reliably estimate the performance of the teams in advance, which can only be achieved by a relatively accurate rating of the clubs. Otherwise, the league phase has a high probability of becoming \emph{unbalanced}, meaning that a particular team mostly plays against opponents with a high number of wins, while another team mainly plays against opponents with a low number of wins. This will certainly be regarded as unfair, analogously to traditional Swiss-system tournaments \citep{Csato2017a}.
The early elimination of strong clubs can also have serious financial consequences as they attract the most attention from both the media and the fans.

\subsection{The implementability of our proposal} \label{Sec52}

This fundamental change in the UEFA Champions League offers a unique opportunity to modify the calculation of the UEFA club coefficient, too.
Two recent reforms, approved by the UEFA and the F\'ed\'eration Internationale de Football Association (International Association Football Federation, FIFA), reinforce that our recommendation has a reasonable chance to be implemented in practice:
\begin{itemize}
\item
UEFA has used the proposal of \citet{Guyon2018a} to minimise group advantage in the 2020 UEFA European Football Championship, making the tournament fairer than its previous edition;
\item
FIFA has developed and introduced a new model for calculating the FIFA World Ranking based on the Elo method in 2018 \citep{FIFA2018c}.
\end{itemize}
The current algorithm of the FIFA World Ranking ``\emph{is not only intuitive, easy to understand and improves overall accuracy of the formula, but also addresses feedback received about the previous model and provides fair and equal opportunities for all teams across all confederations to ascend the FIFA World Ranking}'' \citep{FIFA2018c}. Consequently, using the Elo method to quantify the strength of European football clubs seems to be a promising recommendation in order to guarantee fairer schedules in the novel tournament design of the UEFA Champions League.

\section{Conclusions} \label{Sec6}

This study has investigated the ability of two measures of strength---the official UEFA club coefficient and the alternative Football Club Elo Rating (\url{http://clubelo.com/})---to predict the performance of the teams playing in the UEFA Champions League, a highly prestigious and popular football competition. For the sake of comparability, the Elo ratings have also been fixed at the beginning of each season. Since the club coefficient does not take the games played in the national leagues and cups into account, it is not surprising that it is outperformed by the Elo rating, which contains this information.

Our findings can be interesting for tournament organisers, especially for administrators who are responsible for the methodology of coefficients used for ranking, seeding, and distributing prize money in sports competitions.
In particular, we have a clear message for the Union of European Football Association (UEFA): it is time to reform the calculation of club coefficients used for seeding and distributing prize money \citep{Csato2023f, UEFA2022c} in European club football. This would be especially important because the new tournament format of the Champions League, to be introduced in the 2024/25 season, requires an accurate measurement of teams' strength in order to create a fair schedule.

Naturally, the Football Club Elo Rating is not necessarily the best possible predictor. The Elo algorithm is able to incorporate several characteristics of the matches and preferences of the decision-makers \citep{SzczecinskiRoatis2022}; for instance, games played in European competitions could have a higher weight compared to games played in the national leagues. Hopefully, testing and comparing these variants will be the topic of future papers. In addition, simulations may uncover the sporting effects of using an inaccurate measure of strength such as the current UEFA club coefficient, and the importance of finding the hidden ranking of the teams in various competition formats.

\section*{Acknowledgements}
\addcontentsline{toc}{section}{Acknowledgements}
\noindent
This paper could not have been written without \emph{Gergely Bodn\'ar}, who has helped with data collection. \\
\emph{Kolos Csaba \'Agoston}, \emph{Andr\'as Gyimesi}, and \emph{D\'ora Gr\'eta Petr\'oczy} have provided valuable comments and suggestions. \\
Three anonymous reviewers gave useful remarks on earlier drafts.

\bibliographystyle{apalike}
\bibliography{All_references}

\end{document}